\documentstyle[12pt]{article}

\advance\textwidth 1cm
\advance\oddsidemargin -0.5cm

\newcommand\beq{\begin{equation}}
\newcommand\eeq{\end{equation}}
\def\beqa{\begin{eqnarray}}
\def\eeqa{\end{eqnarray}}
\def\be{\[}
\def\ee{\]}

\def\H#1{{H^{#1}}}

\def\sHa{{\cal H}}

\def\eps{{\epsilon}}
\def\la{{\lambda}}
\def\non{{\nonumber}}

\def\U{{\cal U}}

\def\vb{{\vphantom{b}}}

\def\c#1{{\vphantom{b}_{,#1}}}
\def\d{{\partial}}

\def\ps#1{{\psi^{(#1)}}}

\begin{document}
\title{Quantum Canonical Transformations\\ and Integrability:\\
Beyond Unitary Transformations}
\author{Arlen Anderson\thanks{arley@ic.ac.uk }\\
Blackett Laboratory\\ Imperial College\\ Prince Consort
Road\\ London SW7 2BZ England}
\date{Rev. Aug. 27, 1993}
\maketitle

\vspace{-10cm}
\hfill Imperial/TP/92-93/19

\hfill hep-th/9302061
\vspace{10cm}

\begin{abstract}
Quantum canonical transformations are defined in analogy to classical
canonical transformations as changes of the phase space variables which
preserve the Dirac bracket structure.  In themselves, they are neither
unitary nor non-unitary.  A definition of quantum integrability in terms
of canonical transformations is proposed which includes systems
which have fewer commuting integrals of motion than degrees of freedom.
The important role of non-unitary transformations in integrability is
discussed.
\end{abstract}
\newpage

\baselineskip=24pt

Unitary transformations are a cornerstone of quantum theory. Despite
Dirac's assertion\cite{Dir}, however, they fall short of being the analog
of the classical canonical transformations. Quantum canonical
transformations can be defined without specifying a Hilbert space
structure, and in themselves they are neither unitary nor non-unitary.

Canonical transformations play essentially three distinct roles: in
evolution, in physical equivalence and in integrability. Evolution is
described by unitary canonical transformations---this is the source of the
analogy between unitary and classical canonical transformations\cite{Koo}.
Physically equivalent theories are related by isometric
transformations\cite{And3}, of which the unitary transformations\cite{Mos}
are an important subclass. This Letter will define general quantum
canonical transformations and will illustrate the importance of
non-unitary transformations for quantum integrability.

Using quantum canonical transformations, the wave equation for a system of
interest can be transformed to a simpler equation whose general solution is
known. Because the transformation is defined outside the Hilbert space
structure of the theory, it transforms all solutions of the wave equation,
not just the normalizable ones. As well, the norm of states may not be
preserved by the transformation.

The argument is made below that a general quantum canonical transformation
can be decomposed as a product of elementary canonical transformations of
known behavior, as conjectured by Leyvraz and Seligman\cite{LeS}. Each of
the elementary canonical transformations corresponds to a familiar tool
used in solving differential equations: extracting a function of the
independent variables from the dependent variable, change of independent
variables, and Fourier transform.  The significance of this is that the
procedure of solving a linear differential equation is systematized by the
canonical transformations.  More sophisticated tools, including
raising and lowering operators\cite{InH}, intertwining
operators\cite{And,And2}, and differential realizations of Lie
algebras\cite{Mil}, are easily shown to be canonical transformations in
this sense. Few of these are unitary transformations, yet together they
solve nearly all known integrable models in quantum mechanics.

As this approach takes place outside of a Hilbert space context,
there are two unfamiliar
distinctions that must be made\cite{And4}. First, the non-commuting phase space
variables $(q,p)$ are to be understood not as operators but as elements of
an associative algebra $\U$ generated by complex functions\cite{arb} of
$q,\, p,\, q^{-1},\, p^{-1}$, consistent with the canonical commutation
relations. As elements of this algebra, functions like $p^{-n}$ are
well-defined. The variables $(q,p)$ have a representation as operators
$(\check{q},\check{p})\equiv (q, -i\d_q)$ acting on functions $\psi(q)$
on configuration space. These are not to be thought of as self-adjoint
operators in the
standard inner product because no Hilbert space has been specified (and in
particular the functions $\psi(q)$ need not be square-integrable).

Functions $C(q,p)\in \U$ are represented by operators
$\check{C}(\check{q},\check{p})$. Operators involving $(p^{-1})\check{\vb}$ are
to be understood in the sense of pseudo-differential operators\cite{Hor}.
To avoid technical detail, the domains of operators are not given, but are
to be inferred from their behavior. There is a subtlety in the
correspondence of functions in $\U$ and their representation as operators:
the operator $(C^{-1})\check{\vb}$ corresponding to $C^{-1}$ is not always
inverse to $\check{C}$ because the kernels of $\check{C}$ or
$(C^{-1})\check{\vb}$ may be non-trivial\cite{ker}.
While this prevents one from
rigorously speaking of the operator $\check{C}^{-1}$, except when
$\check{C}$ is invertible, by using $(C^{-1})\check{\vb}$, one
effectively defines the inverse for all functions lying outside the kernels
of the respective operators.

To allow for time-dependent transformations, it is useful to extend the
phase space to include time $q_0$ and its conjugate momentum $p_0$, with
$[q_0,p_0]=i$. For notational convenience, let $(q,p)$ denote all of the
extended phase space variables: equations will be given as if $(q,p)$ were
one-dimensional; the extension to higher dimensions is straightforward.

A classical canonical transformation is a change of the classical phase space
variables $(q_c,p_c)\mapsto (q'_c(q_c,p_c),p'_c(q_c,p_c))$
which preserves the
Poisson bracket $\{q_c,p_c\}=1=\{q'_c,p'_c\}$. A general quantum canonical
transformation may be defined in direct analogy as a change of the
(non-commuting) phase space variables which preserves the Dirac bracket
\beq
\label{Dbr}
[q,p]=i=[q'(q,p),p'(q,p)].
\eeq
These transformations are generated by an arbitrary complex function
$C(q,p)\in \U$
(cf. \cite{Hei})
\beq
CqC^{-1} = q'(q,p) , \quad
CpC^{-1} = p'(q,p).
\eeq
The $C$ producing a given pair $(q',p')$ is unique (up to a
multiplicative constant).
Note that factor ordering is
built into the definition of the canonical transformation in the ordering
of $C$.  No Hilbert space is mentioned in this definition.

The Schrodinger operator corresponds to the function $\sHa(q,p)= p_0 +
H(q_i,p_i,q_0)$ in $\U$. The canonical transformation $C$ transforms this
as
\beq
\label{sHtran}
\sHa'(q,p) = C\sHa(q,p)C^{-1}
= \sHa(Cq C^{-1},Cp C^{-1}).
\eeq
(Generalizing the notion of canonical transformation, one could consider
inhomogeneous
transformations $\sHa'=DC\sHa C^{-1}$, $D\in \U$;  $D=1$ is
assumed here.) Solutions of $\check{\sHa}'\psi'=0$ induce solutions of
$\check{C}\check{\sHa}(C^{-1})\check{\vb}\,\psi'=0$.  If the kernel
of $\check{C}$ is trivial, then
\beq
\label{wftran}
\psi=(C^{-1})\check{\vb}\,\psi'
\eeq
are solutions of $\check{\sHa}$.
Note that since no inner product has been specified, the transformation
$(C^{-1})\check{\vb}$
acts on all solutions of $\check{\sHa'}$, not merely the normalizable
ones.  If $\ker (C^{-1})\check{\vb}$ (or $\ker\check C$) is non-trivial,
then additional
canonical transformations between $\sHa$ and $\sHa'$ may be needed to
construct all the solutions of
$\sHa$.  The uniqueness of the transformation $C$ is discussed below.

When the kernel of $\check{C}$ is non-trivial, the situation is less
simple and requires further discussion.
In this case, there may be solutions $\psi'$ of $\check{\sHa'}$
which by
(\ref{wftran}) produce a $\psi$ which is not a solution of $\check{\sHa}$, but
instead lead to
\beq
\psi''=\check{\sHa}\psi,
\eeq
where
\be
\psi''\in \ker \check{C}.
\ee

To illustrate the problem in a simple case, consider $\sHa=p^3$,
$\sHa'=p^3$.  Clearly, $C=p$ is a canonical transformation, $C\sHa
C^{-1}= \sHa'$.  Consider the solution $\psi'=q^2$ of $\check{\sHa'} \psi'=0$.
By (\ref{wftran}), this gives $\psi=(p^{-1})\check{\vb} \psi'=iq^3/3$.
This is not a solution of $\check{\sHa}$: $\check{\sHa}\psi= -2 \in\ker
\check{C}$.
One has $\check{p} \psi=q^2 =\psi'$ so that $\check{C}$ is invertible on
the solution $\psi$, so this is not the source of the problem.

The problem is
that when $\ker\check{C}$ is non-trivial, the transformation
$(C^{-1})\check{\vb}$ can take one outside the
solution space of $\check{\sHa}$.
To deal with this, one must
always check that $\check{\sHa}\psi=0$ for candidate
$\psi=(C^{-1})\check{\vb} \psi'$.  If $\psi$ is not a solution, it has a
decomposition $\psi=\psi_s+ \psi_n$, as the sum of a solution $\psi_s$ and a
non-solution $\psi_n$.  If the
intersection of $\ker\check{C}$ and $\ker(\sHa^{-1})\check{\vb}$ is
empty, then $\check{\sHa}$ is invertible on $\psi_n$.  Thus, one may remove it
from $\psi$ by the projection
$$\psi_s=(1-(\sHa^{-1})\check{\vb}\check{\sHa})\psi.$$
If  $\ker\check{C} \cap \ker(\sHa^{-1})\check{\vb} \not= \emptyset$, one
must work harder.
A completely general method of handling
the non-trivial kernel of $\check{C}$ is not yet worked out.

Consider the uniqueness of the canonical
transformation $C$ between $\sHa$ and $\sHa'$. A symmetry of $\sHa$ is a
transformation $S_\la$ such that $S_\la \sHa S_\la^{-1}=\sHa$. The
symmetries of $\sHa$ form a group. If $\sHa$ has a symmetry $S_{\la}$ and
$\sHa'$ a symmetry $S'_{\mu}$, then the function $S^{\prime\, -1}_{\mu} C
S_{\la}$ is also a canonical transformation from $\sHa$ to $\sHa'$.
Conversely, if $C_a$ and $C_b$ are two canonical transformations from
$\sHa$ to $\sHa'$, then $C_b^{-1}C_a$ is a symmetry of $\sHa$ and $C_a
C_b^{-1}$ is a symmetry of $\sHa'$. This implies that the collection $\cal
C$ of canonical transformations from $\sHa$ to $\sHa'$ are given by one
transformation $C$ between them and the symmetry groups of $\sHa$ and
$\sHa'$.

An observable (integral of motion) $A$ is a function in $\cal U$ which
commutes with $\sHa$, $[A,\sHa]=0$.
Since canonical transformations preserve the commutation relations, they
induce transformations on the observables of a theory.  The observables $A$
which commute with $\sHa$ are obtained from the $A'$ that commute with
$\sHa'$ by
\beq
A=C^{-1} A' C.
\eeq
The eigenvalues of a complete set of commuting observables are often used to
characterize quantum states.  The observables which characterize the
states of $\check\sHa$ are thus induced from those which characterize the
states of $\check\sHa'$.  Suppose that $\sHa'$ has a complete set of
commuting observables. Then if $\check C$ is invertible, this set
is transformed to a complete set for
$\sHa$.  This is the familiar situation that one is accustomed to call
``integrable'':  $\sHa$ has a complete set of commuting observables.

An unexpected form of integrability is possible in the case that more
than one canonical
transformation is needed to obtain all the solutions of $\check\sHa$.
This may happen if $\check C$ is not invertible.
Suppose that two canonical transformations $C_1$ and $C_2$ suffice to
obtain all the solutions of $\check\sHa$.  By assumption, there are
solutions  $\psi_1= (C_1^{-1})\check{\vb} \psi_1'$ of $\check\sHa$ which
cannot obtained from any solution $\psi_2'$ of $\sHa'$ using
$(C_2^{-1})\check{\vb}$.  For example, the state of $\sHa'$ that
should correspond to $\psi_1$ may lie in $\ker(C_2^{-1})\check{\vb}$,
or possibly $\psi_1\in \ker \check C_2$.  Similarly, there
are solutions $\psi_2= (C_2^{-1})\check{\vb} \psi_2'$ which cannot be
obtained using $(C_1^{-1})\check{\vb}$.  Together, however, all solutions
of $\sHa$ are encompassed by solutions of the form $\psi_1$ and $\psi_2$.

Let $A_k'$ denote a complete set of commuting
observables which characterize the states of $\sHa'$.
Two sets of observables $A_{1k}$ and
$A_{2k}$ are obtained from $A_k'$,
\beq
A_{1j}=C_1 A_j' C_1^{-1}, \quad A_{2k}=C_2 A_k' C_2^{-1}.
\eeq
In general, $A_{1j}$ and $A_{2k}$ will not commute for all $j,k$.  The
result is
that while all the states of $\sHa$ have been constructed from those
of $\sHa'$, the states of $\sHa$ are not characterized by a single
complete set of commuting
observables.  Instead, the states $\psi_1$ and $\psi_2$ are
characterized by different sets of commuting observables.  This
is a more general form of integrability than has been traditionally
considered.  In a system of this kind having $n$ degrees of freedom, when
considering all states of the system, there would
appear to be fewer than $n$ quantum integrals of the motion (observables).
On suitably restricted subsets of states, however,
there would be different collections of $n$ integrals.

Consider now the construction of canonical transformations.
Classically, the infinitesimal generating functional
$F(q_c,p_c)$ generates the finite canonical transform\-ation\cite{Gol}
\beq
u'(q_c,p_c)=\exp(\eps v_F) u(q_c,p_c).
\eeq
where $v_F=F\c{p_c} \d_{q_c}- F\c{q_c} \d_{p_c}$ is the Hamiltonian vector
field
generated by $F$.  The algebra of the canonical group is
\beq
[v_F,v_G]=-v_{\{F,G\} },
\eeq
and it is generated by the Hamiltonian vector fields obtained from
$F\in \{h(q_c),\ h(q_c)p_c,\newline h(p_c),\ h(p_c)q_c \}$.
In principle then a general classical canonical transformation can be
expressed as a product of finite transformations with these $v_F$.

Quantum mechanically, each of these classical transformations has a
quantum implementation as $C=e^{iF}$ (note that $F$ is in general complex).
Introducing the operation $I$ which interchanges the coordinate and
momentum, $(q,p)\mapsto (-p,q)$, the transformations which are nonlinear
in the momentum can be expressed in terms of the other two.
There are then three elementary canonical transformations(cf. \cite{Dee}):\\
1) similarity (gauge) transformations, $C=e^{-f(q)}$
\beq
\label{sim}
(q,p) \mapsto (q,p-if\c{q}),\quad \psi'(q)=e^{-f(q)}\psi(q)
\eeq
2) point canonical transformations, $C=P_{f(q)}$
\beq
(q,p) \mapsto (f(q), {1\over f\c{q}}p),\quad \psi'(q)=\psi(f(q))
\eeq
and, 3) interchange, $C=I=(2\pi)^{-1/2} \int_{-\infty}^\infty dq e^{iqq'}$
\beq
\label{int}
(q,p) \mapsto (-p',q'),\quad \psi'(q')=I\psi(q).
\eeq
The point canonical transformation is denoted by $P_{f(q)}$
because in general a finite product of terms of the form
$\exp(ig(q)p)$ is required to represent such a transformation\cite{Mil2}.
The transformations non-linear in the momentum are the composite
elementary transformations:\\
4) $C=e^{-f(p)}=I e^{-f(q)} I^{-1}$
\beq
(q,p) \mapsto (q+if\c{p},p), \quad \psi'(q)=e^{-f(\check{p})} \psi(q)
\eeq
and, 5) $C=P_{f(p)}=I P_{f(q)} I^{-1}$
\beqa
(q,p) &\mapsto& ({1\over f\c{p}}q, f(p)),\\
&\quad&\hspace{-0.5cm} \psi'(q')= P_{f(p)} \psi(q)= (f^{-1}(p)\c{p})\check{\vb}
\exp(if^{-1}(\check{p})q') \psi(q) |_{q=0}. \non
\eeqa
It is important to emphasize that all functions are complex and may have
zeroes or singularities. All expressions are ordered as written. The
functions in the transformations may be many-variable. Since coordinates
and momenta of differing index commute, a variable participates only as a
constant parameter in any transformation which does not involve its
conjugate.

Since a general classical canonical transformation can be expressed as a
product of elementary canonical transformations, one expects that the same
is true for quantum canonical transformations.
This is equivalent to the assertion that any
function $C(q,p)$ can be decomposed as a product of the elementary
canonical transformations. There are functions $C$ which cannot decomposed
into a finite product, and their action on a wavefunction cannot be realized
explicitly.

This motivates the following definition of quantum
integrability:
\begin{quote}
Definition. A quantum system $\sHa(q,p)$ is {\em integrable} (in the sense of
homogeneous canonical transformations) if its general solution $\psi$ can
be obtained from arbitrary time-independent functions $\ps0$ using
a collection of finitely decomposable canonical
transformations $C_\la\in \cal C$ which trivialize the wave operator
\beq
\label{qint}
C_\la\sHa(q,p)C_\la^{-1}=p_0.
\eeq
\end{quote}
Note that the $C_\la$ can be expressed as $C S_\la$ where $C$ is a particular
canonical transformation to triviality and $S_\la$ is a symmetry of $\sHa$.
If $A_k'$ are a complete set of commuting observables for $\sHa'=p_0$,
then $A_{k\la}=C_\la A_k' C_\la^{-1}$ are sets of commuting observables for
$\sHa$ for each $\la$.  The $A_{k\la}$ may not commute for different $\la$.
As discussed above, the system is nevertheless
integrable, even though there is not a single set of commuting observables
which serves to characterize all states of $\sHa$.

The condition of finite decomposability is necessary to have explicit
representations for the solutions of $\sHa$.  It raises the question of
characterizing the class of Hamiltonians that can solved with a finite
number of elementary transformations.  This is reminiscent of the basic
question addressed by
Galois theory of which polynomials can be factored using a
finite combination of the operations of addition, subtraction,
multiplication, division and the taking of $n^{\rm th}$ roots.  Just as
there are polynomials whose roots cannot be expressed in terms of a finite
combination of the algebraic operations, one expects there are equations
which cannot be solved by a finite number of elementary canonical
transformations.

Having established the basic formalism, consider some illustrative
examples\cite{And4}.  The time-independent Schrodinger equation,
with
$\sHa=p_0+H(q_i,p_i)$, is clearly trivialized by $C=e^{iH(q_0-t)}$
(where $t$ is a constant).  In general, this is not an elementary
transformation, and its action on the wavefunction is not immediately
evident.  By finding a (finitely decomposable)
canonical transformation $\tilde C$ such that, say, $\tilde C H \tilde
C^{-1}=p$, the action of $C$ is determined because
\beq
e^{iH(q_0-t)}=\tilde C^{-1} e^{ip(q_0-t)} \tilde C
\eeq
is now a finite product of elementary canonical transformations.
Applying the operator representation of this to $\delta(q-q')$,
one can compute the propagator $K(q,q_0|q',t)$.

This procedure may be used in general to simplify functions of
operators.  Consider the one-dimensional point canonical
transformation $e^{iag(q)p}$.
Let $G(q)=\int dq/g(q)$.  For $C=P_{G(q)}$, one has
$Cp C^{-1}= g(q)p$.  The
action of $e^{iag(q)p}$ on $q$ is then computed
\beqa
\exp(iag(q)p) q \exp(-iag(q)p) &=& Ce^{iap} C^{-1} q C e^{-iap} C^{-1}
\non \\
&=& G^{-1}(G(q)+a) .
\eeqa
This result is found by a more laborious method in \cite{Dee}.

Canonical transformations involving polynomial functions of $p$ are
non-unitary in inner products with coordinate-valued measure
density\cite{And3}. As they underlie raising and lowering operators, the
recursion operators for the special functions, intertwining and Lie
algebraic transformations, they are undeniably important in the solution of
many problems. As an illustration, consider the Darboux
transformation\cite{And2,Dar} from a Hamiltonian $H_0=p^2 +V_0$ to another
$\H1=p^2+V_0 -2 g\c{q}$, where $g$ satisfies the Ricatti equation
$g\c{q}+g^2 =V_0+\la$. The canonical transformation from $H_0$ to $H_1$ is
$C=\exp(\int g dq)p \exp(-\int g dq)$. The key step in the transformation
is that performed by $p$ which transforms
\beq
\label{dio}
q \mapsto q -{i\over p}= p q {1\over p}.
\eeq
This has the remarkable property
\beq
g(q - {i\over p}) = pg(q){1\over p}
=g(q)-ig(q)\c{q}{1\over p }.
\eeq
The Taylor expansion of $g$ terminates at the first term; classically,
there would be an infinite series.

{}From (\ref{sim})-(\ref{int}), it is clear as discussed in the introduction
that the elementary canonical transformations correspond to the standard
tools used in the solution of differential equations. The discovery and
implementation of transformations to solve an equation is made
more transparent when looked at
from the perspective of canonical transformations. The practical gain is
largely through a reduction in the technical demands of implementing a
trial transformation.

The integrability of
a quantum system by (\ref{qint}) corresponds to the existence of a sequence
of standard manipulations which solve the wave equation.  This notion of
integrability is different than the standard one of the existence of a
complete set of commuting observables. The possibility exists that
all the solutions of a Hamiltonian can be found using canonical
transformations, but there will not be a single complete set of commuting
observables valid for all states.  Rather there will be
collections of commuting
observables which apply to different sets of states.

Acknowledgement.  I would like to thank C.J. Isham for discussions of
this work.


\begin{thebibliography}{99}
\bibitem{arb} The largest class of admissible functions and its topology
for series, e.g. $(1+q)^{-1}$, is not yet known
precisely. It must include all functions having Laurent expansions and
their algebraic and functional inverses, e.g. $q^{1/2}$, $\ln q$,
but probably excludes distributions, e.g. $\theta(q)$, $\delta(q)$.
\bibitem{ker} A function $\psi$ is said to be in $\ker \check C$ if $\check C
\psi$ either vanishes or diverges.
\end{thebibliography}
\end{document}